\documentclass{moriond}

\def\be{\begin{equation}}
\def\ee{\end{equation}}
\def\bea{\begin{eqnarray}}
\def\eea{\end{eqnarray}}

\newcommand{\Photo}{\includegraphics[height=35mm]{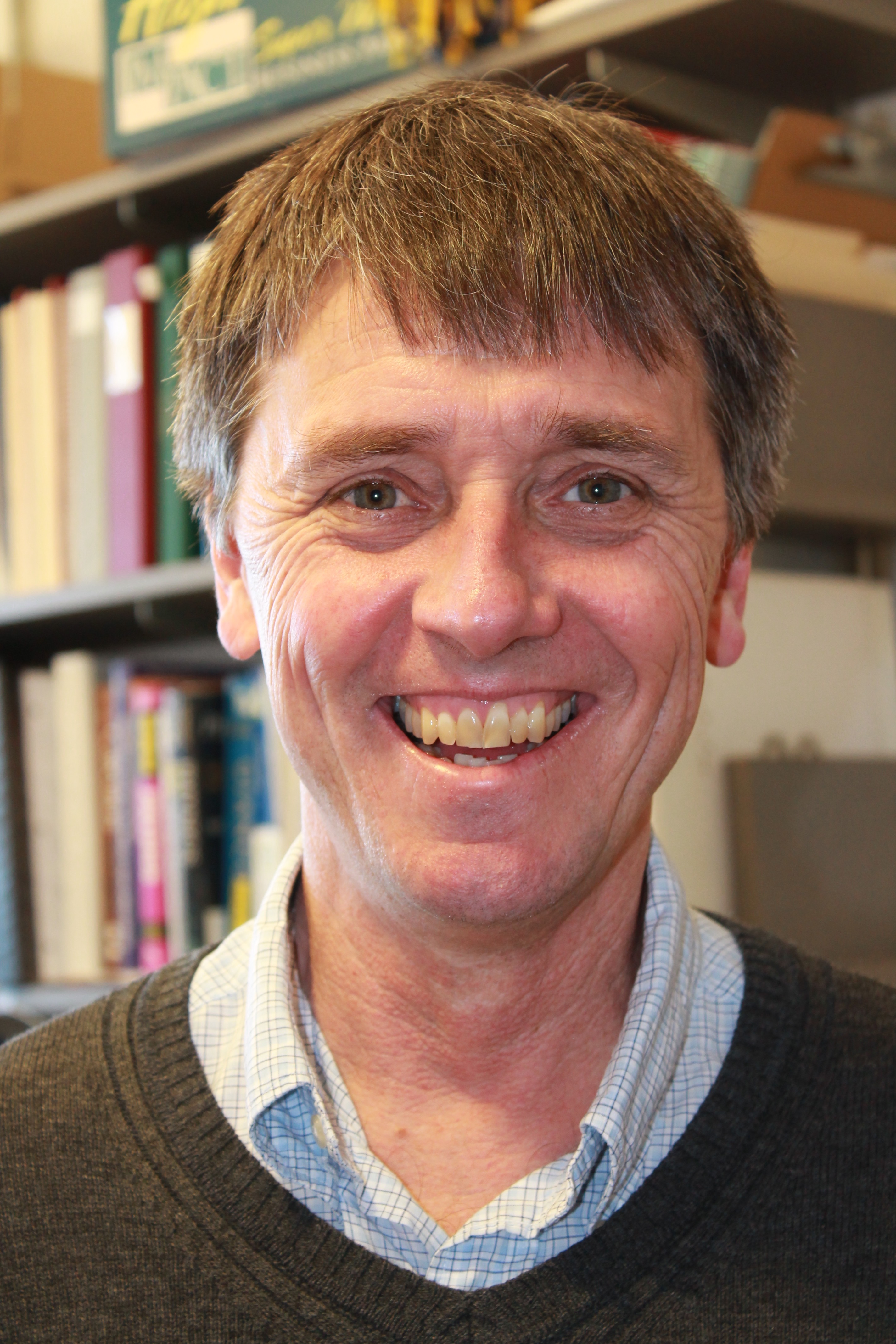}}

\begin{document}
\vspace*{4cm}
\title{Implications of the binary coalescence events found in O1 and O2 for the stochastic background of gravitational events}

\author{N.L. Christensen \\
(For the LIGO Scientific Collaboration and the Virgo Collaboration)}

\address{ARTEMIS, Universit\'e C\^ote d'Azur, Observatoire C\^ote d'Azur, CNRS, 06304 Nice, France \\
Department of Physics, Carleton College, Northfield, Minnesota 55057, USA}

\maketitle\abstracts{
The Advanced LIGO and Advanced Virgo detectors have commenced observations. Gravitational waves from the merger of binary black hole systems and a binary neutron star system have been observed. A major goal for LIGO and Virgo is to detect or set limits on a stochastic background of gravitational waves. A stochastic background of gravitational waves is expected to arise from a superposition of a large number of unresolved cosmological and/or astrophysical sources. A cosmologically produced background would carry unique signatures from the earliest epochs in the evolution of the Universe. Similarly, an astrophysical background would provide information about the astrophysical sources that generated it. The observation of gravitational waves from compact binary mergers implies that there will be a stochastic background from these sources that could be observed by Advanced LIGO and Advanced Virgo in the coming years. The LIGO and Virgo search for a stochastic background should probe interesting regions of the parameter space for numerous astrophysical and cosmological models. Presented here is an outline of LIGO and Virgo's search strategies for a stochastic background of gravitational waves, including the search for gravitational wave polarizations outside of what is predicted from general relativity. Also discussed is how global electromagnetic noise (from the Schumann resonances) could affect this search; possible strategies to monitor and subtract this potential source of correlated noise in a the global detector network are explained. The results from Advanced LIGO's observing runs O1 and O2 will be presented, along with the implications of the gravitational wave detections. The future goals for Advanced LIGO and Advanced Virgo will be explained. }

\section{Introduction}
A consequence of Einstein's general theory of relativity are gravitational waves, perturbations to spacetime that travel away from their source at the speed of light. 
A stochastic gravitational-wave background (SGWB) signal is formed from the superposition of many events or processes that are too weak and too numerous to be resolved individually, and which combine to produce a SGWB~\cite{Christensen_2019}. 
Cosmological sources, such as inflation, pre-Big Bang models, or cosmic strings, could create a SGWB.
Astrophysical sources can also create a SGWB; this background could be produced over the history of the Universe from compact binary coalescences, supernovae, and neutron stars. 
In fact, the recent observations by Advanced LIGO and Advanced Virgo of gravitational waves in observing runs O1 and O2 from 10 binary black hole mergers~\cite{PhysRevLett.116.061102,PhysRevLett.116.241103,PhysRevX.6.041015,LIGOScientific:2018mvr} and a binary neutron star merger~\cite{PhysRevLett.119.161101} implies that a SGWB will be created from these events happening throughout the history of the universe and it may be detectable by Advanced LIGO~\cite{0264-9381-32-7-074001} and Advanced Virgo~\cite{0264-9381-32-2-024001} in the coming years~\cite{PhysRevLett.116.131102,PhysRevLett.120.091101,LIGOScientific:2019vic}.
As Advanced LIGO and Advanced Virgo conduct their observations a major goal will be to measure the SGWB. 

The spectrum of a SGWB is usually described
by the dimensionless quantity $\Omega_{gw}( f )$ which is the
gravitational-wave energy density per unit logarithmic frequency,
divided by the critical energy density $\rho_{c}$ ($\rho_{c}=3 c^2 H^2_0 /8 \pi G$, where $H_0$ is the present value of the Hubble constant) to close the
universe, 
\begin{equation}
\Omega_{gw}( f ) = \frac{f}{\rho_c} \frac{d \rho_{gw}}{d f} ~.
\end{equation}
Many theoretical models of the SGWB in the observation band of LIGO and Virgo are characterized by a
power-law spectrum which assumes that the fractional energy density in gravitational waves has the form
\begin{equation}
\Omega_{gw}( f ) = \Omega_{\alpha} \left(\frac{f}{f_{ref}}\right)^{\alpha} ~,
\end{equation}
where $\alpha$ is the spectral index and $f_{ref}$ is a reference
frequency. 
Cosmologically produced SGWBs are typically approximated by a power law in the LIGO-Virgo frequency band,  $\alpha = 0$, while $\alpha = 3$ is characteristic of some astrophysical models (and also a flat strain power spectral density spectrum).
A SGWB from binary black holes in Advanced LIGO and Advanced Virgo's most sensitive frequency band ($10$ Hz - $100$ Hz) would have $\alpha = 2/3$.

The method by which LIGO and Virgo have attempted to measure the SGWB is, in principle, not difficult; optimally filtered correlations from the output strain data from two detectors are calculated~\cite{Christensen:1992,stoch_allenromano}. 
This method was used on initial LIGO~\cite{LIGO} and initial Virgo~\cite{Virgo} data to set limits on the energy density of the SGWB~\cite{S5stoch,S5H1H2,S6stoch}.
No signal was detected, but the results constrain the energy density of the SGWB to be $\Omega_0 < 5.6 \times 10^{-6}$ at 95\% confidence~\cite{S6stoch} in the 41.5--169.25 Hz band.
The advanced detectors will ultimately have about 10-times better strain sensitivity than the initial detectors; the low frequency limit of the sensitive band is also extended from 40 Hz down to 10 Hz. As will be described below, this has improved the upper limits on the energy density of the SGWB by a factor of about 100.
Furthermore, the number of detectors operating in a worldwide network will increase, eventually including sites at LIGO-Hanford, LIGO-Livingston, Virgo, GEO-HF (at high frequencies)~\cite{0264-9381-31-22-224002}, KAGRA (Japan)~\cite{PhysRevD.88.043007}, and LIGO-India~\cite{doi:10.1142/S0218271813410101}.
The significant strain sensitivity improvements and wider bandwidth will enable real breakthroughs in the searches for the SGWB, with a potential sensitivity approaching $\Omega_0 \sim 10^{-9}$.
The detection of a cosmologically produced SGWB would be a landmark discovery of enormous importance to the larger physics and astronomy community. The detection of an astrophysically produced SGWB would also be of great interest; the loudest contributions to such an SGWB would likely be stellar mass binary black hole systems and binary neutron star systems, due to their large apparent abundances~\cite{PhysRevLett.116.131102,PhysRevLett.120.091101,LIGOScientific:2019vic}.

Gravitational-wave signals that are too weak to be detected individually combine to form a SGWB.
The SGWB that LIGO and Virgo hope to observe could be created from two classes of sources.
A cosmologically produced SGWB would be created in the earliest moments of the Universe. There are a host of cosmological processes that could contribute to the SGWB, such as the amplification of vacuum fluctuations following inflation~\cite{kolbturner}, phase transitions in the early universe~\cite{starobinksii,bar-kana}, cosmic strings~\cite{kibble,damour,olmez1,olmez2}, and pre-Big Bang models~\cite{buonanno,mandic}.
An astrophysically produced SGWB would arise from the ensemble of what would be considered to be standard astrophysical events~\cite{2011RAA....11..369R}.
In total the astrophysical background would be the result of a broad spectrum of events, including core collapses to neutron stars or black holes~\cite{2005PhRvD..72h4001B,2006PhRvD..73j4024S,2009MNRAS.398..293M,2010MNRAS.409L.132Z,PhysRevLett.116.131102}, rotating neutron stars~\cite{2001A&A...376..381R,2012PhRvD..86j4007R} including magnetars~\cite{2006A&A...447....1R,2011MNRAS.410.2123H,2011MNRAS.411.2549M,2013PhRvD..87d2002W}, phase transitions \cite{sigl,2009GReGr..41.1389D} or initial instabilities in young neutron stars~\cite{1999MNRAS.303..258F,2011ApJ...729...59Z,2004MNRAS.351.1237H,2011ApJ...729...59Z}, compact binary mergers~\cite{farmer,2011ApJ...739...86Z,2011PhRvD..84h4004R,2011PhRvD..84l4037M,2012PhRvD..85j4024W,2013MNRAS.431..882Z} and compact objects around super-massive black holes~\cite{barack,sigl2}. As LIGO and Virgo observe in the advanced detector era, the cosmologically produced SGWB and the astrophysically produced SGWB are both exciting targets for observation.

\section{Results from Advanced LIGO Observing Runs O1 and O2}\label{sec:O1_results}
Advanced LIGO's first observing run O1 went from September 2015 to January 2016, while its second observing run O2 went from November 2016 to August 2017. Advanced Virgo participated in O2 for the month of August 2017. Together LIGO and Virgo detected gravitational waves from 10 binary black hole mergers and a binary neutron star merger~\cite{LIGOScientific:2018mvr}. The data from the two Advanced LIGO detectors, LIGO Hanford and LIGO Livingston, were used for the search for a SGWB. Data quality cuts removed problematic times and frequencies from the analysis. In total for O1, 30 days of coincident data were analyzed, while for O2 the data amounted to 99 days. No SGWB was detected. 

\subsection{Combined O1 and O2 Isotropic Results}
Assuming that the frequency dependence of the energy density of the SGWB is flat, namely $\alpha = 0$, the constraint on the energy density is $\Omega(f) < 6.0 \times 10^{-8}$ with 95\% confidence within the 20 Hz - 86 Hz frequency band~\cite{LIGOScientific:2019vic}. This is a factor of 2.8 better than the upper limit set by using just the O1 data~\cite{PhysRevLett.118.121101}. For a spectral index of $\alpha = 2/3$ the constraint on the energy density is $\Omega(f) < 4.8 \times 10^{-8}$, 
while for $\alpha = 3$ it is $\Omega(f) < 7.9 \times 10^{-9}$
~\cite{LIGOScientific:2019vic} (both with with 95\% confidence, and a reference frequency of $f_{ref} = 25$ Hz when $\alpha \neq 0$). A prior that is flat in $\Omega_{gw}$ has been used.
The O1 and O1 + O2 results have been used to limit cosmic string parameters~\cite{Abbott:2017mem,LIGOScientific:2019vic}, similar to what was done with initial LIGO and initial Virgo~\cite{S5stoch,S5strings}.

The dramatic improvement in the upper limit on the SGWB energy density was important, but not the most important SGWB outcome of observing runs O1 and O2. The observation of the gravitational waves from stellar mass binary black hole mergers~\cite{PhysRevLett.116.061102,PhysRevLett.116.241103,PhysRevX.6.041015,LIGOScientific:2018mvr} and a binary neutron star merger~\cite{PhysRevLett.119.161101} implies that these events are far more numerous in the universe than originally expected. In fact, it is likely that the SGWB produced from these type of events will be at the level of $\Omega_{gw} \sim 10^{-9}$ in the observing band of Advanced LIGO and Advanced Virgo~\cite{PhysRevLett.116.131102,PhysRevLett.118.121101,PhysRevLett.120.091101,LIGOScientific:2019vic}. 

\begin{figure}
\begin{center}
\includegraphics[width=5.5in]{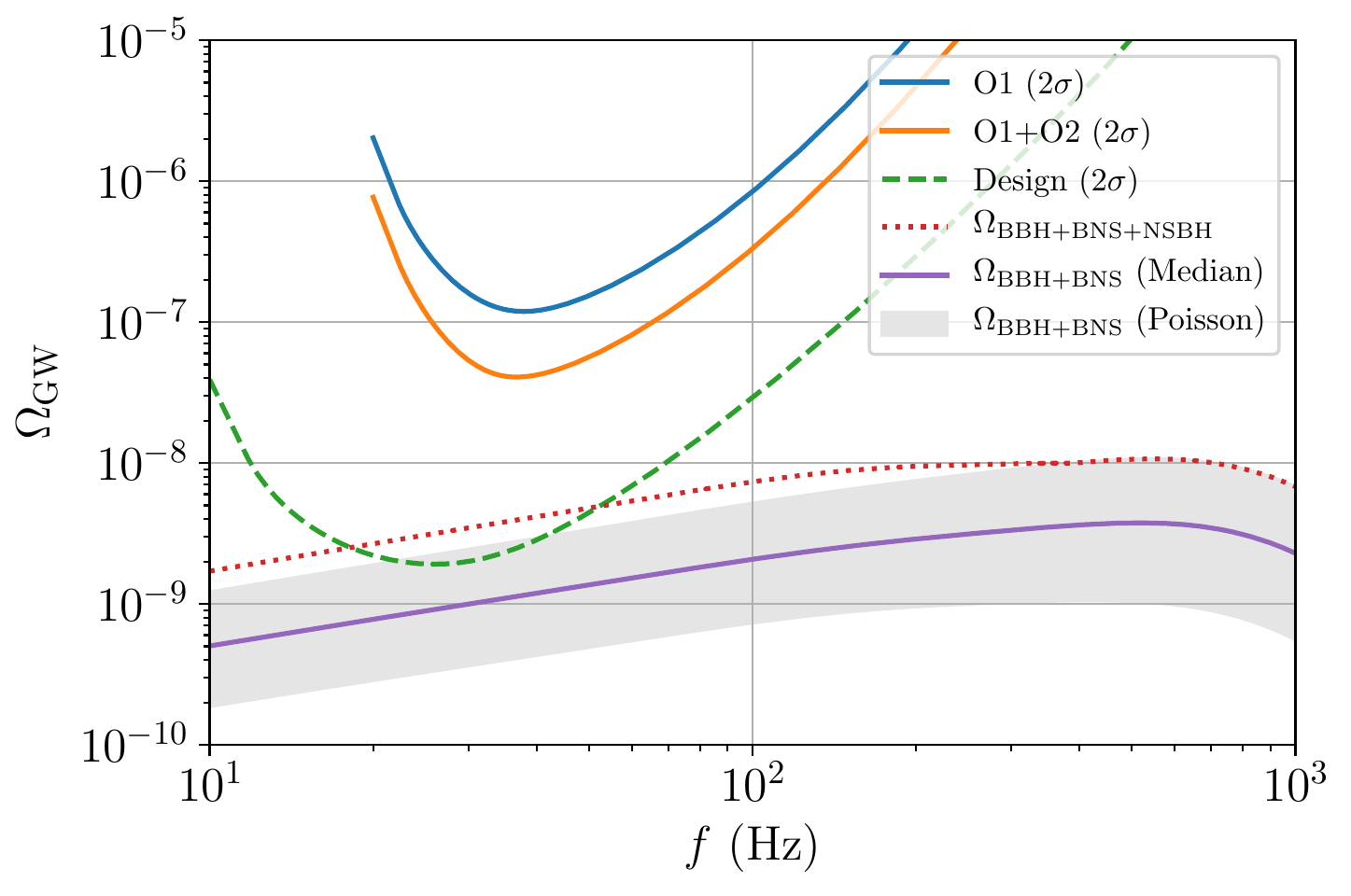}
\caption{The Advanced LIGO SGWB sensitivity curves for O1 \protect \cite{PhysRevLett.118.121101}, combined O1+O2 \protect \cite{LIGOScientific:2019vic}, and design sensitivity \protect \cite{PhysRevD.88.124032,Barsotti_2018}. The purple line is the median total SGWB, combining binary black holes (BBH) and binary neutron stars (BNS); this uses the model presented in \protect \cite{PhysRevLett.120.091101} with updated mass distributions and rates from \protect \cite{LIGOScientific:2018mvr,LIGOScientific:2018jsj}; the gray box is the Poisson error region. The dotted gray line is the sum of the upper limit for the combined BBH and BNS backgrounds with the upper limit on the neutron star - black hole binary (NSBH) background.}
\label{fig:BBH_background}
\end{center}
\end{figure}

Figure~\ref{fig:BBH_background} displays the prediction of the astrophysical SGWB from binary black holes and binary neutron stars, along with the statistical Poisson uncertainties derived from the local binary merger rate.
Also included is the estimate of the contribution from the addition neutron star - black hole binaries.
The same binary formation and evolution scenario is used to compute the SGWB from from neutron star - black hole binaries as in~\cite{PhysRevLett.120.091101}, but an update was made for the mass distributions and rates so as to be consistent with the most recent results given in~\cite{LIGOScientific:2018mvr,LIGOScientific:2018jsj}. For the neutron star - black hole binaries, the same evolution with redshift was used as for the binary neutron stars. 

\subsection{Anisotropic O1 and O2 Results}
Within the LIGO-Virgo observational band it is expected that the SGWB will be essentially isotropic, with the level of assumed anisotropies being many orders of magnitude below the predicted sensitivities~\cite{LIGOScientific:2019gaw}. However, LIGO and Virgo have decided to look for a SGWB that would be anisotropic. Such an anisotropic background could provide even more information about the early universe, or the astrophysical environment in our region of the universe. Using the recent O1 and O2 data there have been three different types of searches for an anisotropic background~\cite{PhysRevLett.118.121102,LIGOScientific:2019gaw}. To look for extended sources, LIGO and Virgo use what is known as the spherical harmonic decomposition~\cite{S5direct}. In order to search for point sources, a broadband radiometer analysis is used~\cite{0264-9381-23-8-S23,S4radiom}. Finally, LIGO and Virgo employed a narrow-band radiometer search to look for gravitational waves in the direction of interesting objects in the sky, such as the galactic center, Scorpius X-1 and SN 1987A.

An anisotropic SGWB was not observed with the Advanced LIGO O1 and O2 data, but important upper limits were set~\cite{PhysRevLett.118.121102,LIGOScientific:2019gaw}.
For broadband point sources, the gravitational wave energy flux per unit frequency was constrained to be  
$F_{\alpha,\Theta}  < (0.05 - 25)\times 10^{-8}$ erg cm$^{-2}$ s$^{-1}$ Hz$^{-1} (f/25$ Hz)$^{\alpha-1}$
depending on the sky location 
$\Theta$
and the spectral power index $\alpha$.  
For extended sources, the upper limits on the fractional gravitational wave energy density required to close the Universe are 
$\Omega(f,\Theta) < (0.19 - 2.89)\times 10^{-8}$ sr$^{-1} (f/25 $ Hz$)^{\alpha}$,
again depending on 
$\Theta$
and $\alpha$.
The directed searches for narrow-band gravitational waves from Scorpius X-1, Supernova 1987~A, and the Galactic Center had median frequency-dependent limits on strain amplitude of 
$h_0 < (4.2,\, 3.6,$ and $ 4.7)\times10^{-25}$
respectively, for the most sensitive detector frequencies 
130 - 175 Hz. See \cite{LIGOScientific:2019gaw} for further details.

\subsection{Tests of General Relativity with the Stochastic Gravitational-Wave Background}
LIGO and Virgo have used the recent observation of gravitational waves from binary black hole and binary neutron star coalescences to test general relativity~\cite{PhysRevLett.116.221101,PhysRevX.6.041015,LIGOScientific:2019fpa,Abbott:2018lct}. The LIGO-Virgo SGWB search has also be extended in order to test general relativity~\cite{Callister:2017ocg}. There is not necessarily a reason to expect extra polarizations of gravitational waves, nor extra polarizations in the SGWB (the consequences would be huge even if the chances are, a priori, not large); however, LIGO and Virgo have the ability to search for these modes, and will do so. With general relativity there are only two possible polarizations for gravitational waves, namely the two tensor modes. Alternative theories of gravity can also generate gravitational waves with scalar or vector polarizations~\cite{Will2014}.

Since there are six possible polarization modes, Advanced LIGO (with only two detectors, that are essentially co-aligned with respect to each other) cannot identify the  polarization of short duration gravitational wave signals~\cite{PhysRevX.6.041015,Romano2017,Will2014}, such as those that have been recently observed~\cite{PhysRevLett.116.061102,PhysRevLett.116.241103,PhysRevX.6.041015}.
A minimum of six detectors would be necessary to resolve the polarization content (scalar, vector and tensor)  of
a short duration gravitational wave~\cite{Will2014}.
A search for long duration gravitational waves, such as those from rotating neutron stars or the SGWB by the two Advanced LIGO detectors,
can directly measure the polarizations of the gravitational waves~\cite{Romano2017,0264-9381-32-24-243001,Isi:2017equ,Callister:2017ocg}.
A detection of a SGWB by Advanced LIGO and Advanced Virgo would allow for a
verification of general relativity that is not possible with short duration gravitational wave searches.

The LIGO-Virgo search for a SGWB has been expanded to a search for 6 polarizations: two tensor modes, two vector modes, and two scalar modes~\cite{Callister:2017ocg}, and applied to the Advanced LIGO Observing Run O1 and O2 data~\cite{PhysRevLett.120.201102,LIGOScientific:2019vic}.
The addition of Advanced Virgo to the network does not improve detection prospects (because of its longer distance displacement from the LIGO detectors), however it will improve the ability to estimate
the parameters of a SGWB of mixed polarizations. The eventual inclusion of KAGRA~\cite{PhysRevD.88.043007} and LIGO-India~\cite{doi:10.1142/S0218271813410101} will further expand the ability to resolve different polarizations of the SGWB, and further test general relativity. Bayesian parameter estimation techniques have been developed in order to search for tensor, vector and scalar polarizations in the LIGO-Virgo data~\cite{Callister:2017ocg}.

When calculating the SGWB for tensor, vector and scalar polarizations one uses the notation $\Omega^A_{\rm gw}(f)$ in analogy with the standard general relativity search. However,  in a general modification of gravity, the quantities $\Omega^T_\mathrm{gw}(f)$, $\Omega^V_\mathrm{gw}(f)$, and $\Omega^S_\mathrm{gw}(f)$ are more correctly associated to measurements of the two-point correlation statistics of different components of the SGWB as opposed to energy densities~\cite{Isi:2018miq}. There was no evidence found for a SGWB of tensor, vector or scalar polarizations in the O1 and O2 data, and upper limits were set~\cite{PhysRevLett.120.201102,LIGOScientific:2019vic}. These upper limits depend on the assumptions made for the what combination of the three polarizations are presumed to be present in the data. For example, when assuming that the tensor, vector or scalar polarizations could all be present in the data, then the upper limits are presented in Table~\ref{tab:nongrUL} assuming a prior that is flat in $\Omega_{gw}$, and another prior that is flat in log$\Omega_{gw}$.

\begin{table}
\begin{center}
\setlength{\tabcolsep}{10pt}
\renewcommand{\arraystretch}{1.2}
\begin{tabular}{l|c c}
\hline
\hline
Polarization &
Uniform prior &
Log-uniform prior  \\
\hline
Tensor $\Omega^T_\mathrm{gw}(f)$ & $8.2 \times 10^{-8}$ & $3.2 \times 10^{-8}$ \nonumber \\
Vector $\Omega^V_\mathrm{gw}(f)$ & $1.2 \times 10^{-7}$ & $2.9 \times 10^{-7}$ \nonumber \\
Scalar $\Omega^S_\mathrm{gw}(f)$ & $4.2 \times 10^{-7}$ & $6.1 \times 10^{-7}$ \nonumber \\
\hline
\hline
\end{tabular}
\caption{Upper limits for the SGWB assumed to be made up of simultaneous contributions of tensor, vector and scalar polarizations: $\Omega^T_\mathrm{gw}(f)$, $\Omega^V_\mathrm{gw}(f)$, and $\Omega^S_\mathrm{gw}(f)$. Results are presented using a prior that is log uniform, and another that is uniform on the amplitude $\Omega_{\rm ref}$ for each polarization. The O1 and O2 data are used. See~ \protect \cite{LIGOScientific:2019vic} for more details.}
\label{tab:nongrUL}
\end{center}
\end{table}

\section{Correlated magnetic noise in global networks of gravitational-wave detectors}
A search for the SGWB uses a cross-correlation between the data from two detectors. Inherent in such an analysis is the assumption that the noise in one detector is statistically independent from the noise in the other detector. Correlated noise would introduce an inherent bias in the analysis. It is for this reason that the data from two separated detectors is used. At one time initial LIGO had two co-located detectors at the LIGO Hanford site. An attempt was made to measure the SGWB with these two detectors, but correlated noise at low frequencies contaminated the measurement, and a clean analysis could only be made above 460 Hz~\cite{S5H1H2}.

The LIGO, Virgo and KAGRA detector sites are thousands of kilometers from one another, and the simple assumption is that the noise in the detectors at these sites is independent from one another. However, this assumption has been demonstrated to be false for magnetic noise. The Earth's surface and the ionosphere act like mirrors and form a spherical cavity for extremely low frequency electromagnetic waves. The Schumann resonances are a result of this spherical cavity, and resonances are observed at  8, 14, 20, 26, ... Hz~\cite{Sentman}. Most of these frequencies fall in the important SGWB detection band (10 Hz to 100 Hz) for Advanced LIGO and Advanced Virgo. The resonances are driven by the 100 or so lightning strikes per second around the world.
The resonances result in magnetic fields of order 0.5 - 1.0 pT Hz$^{1/2}$ on the Earth's surface~\cite{Sentman}. 
In the time domain, 10 pT bursts appear
above a 1 pT background  at a rate of $\approx$ 0.5 Hz ~\cite{FULLEKRUG1995479}.

This magnetic field noise correlation has been observed between magnetometers at the LIGO, Virgo and KAGRA sites~\cite{Schumann,PhysRevD.97.102007}. Magnetic fields can couple into the gravitational wave detectors and create noise in the detectors' output strain channel. It has been determined that the correlated magnetic field noise did not affect the SGWB upper limits measured by initial LIGO and Virgo. For the observing runs O1 and O2 it has been demonstrated that the upper limits on the SGWB were not contaminated by correlated magnetic noise~\cite{LIGOScientific:2019vic}. However, it is possible that correlated magnetic noise could contaminate the future results of Advanced LIGO and Advanced Virgo~\cite{wiener}. If that is the case, then methods must be taken to try and monitor the magnetic fields and subtract their effects. This could be done, for example, via Wiener filtering~\cite{wiener,0264-9381-33-22-224003,PhysRevD.97.102007}. Low noise magnetometers are now installed at the LIGO and Virgo sites in order to monitor this correlated magnetic noise, and to be used if Wiener filtering is necessary for the SGWB searches. In addition to long term magnetic noise correlations, short duration magnetic transients, produced from lightning strikes around the world, are seen to be coincidently visible at the detector sites and could affect the search for short duration gravitational wave events~\cite{0264-9381-34-7-074002}.

\section{Future Observing Runs for LIGO and Virgo}
Advanced LIGO and Advanced Virgo have completed two observing runs, and the results of the search for a SGWB have been published~\cite{PhysRevLett.118.121101,PhysRevLett.118.121102,LIGOScientific:2019vic,LIGOScientific:2019gaw}. At the time of this writing Advanced LIGO and Advanced Virgo are in the middle of the third observing run O3. Over the next few years further observing runs will happen as Advanced LIGO and Advanced Virgo approach their target sensitivities~\cite{Aasi:2013wya}. At their target sensitivities LIGO and Virgo should be able to constrain the energy density of the SGWB to approximately $\Omega_{gw} \sim 2 \times 10^{-9}$ with two years of coincident data (in the 10 Hz to 100 Hz band). At this point LIGO and Virgo could possibly observe a binary black hole and binary neutron star produced SGWB~\cite{PhysRevLett.118.121101,PhysRevLett.116.131102,PhysRevLett.120.091101,LIGOScientific:2019vic}. Various cosmological models~\cite{starobinksii,bar-kana,buonanno,mandic}, or cosmic strings~\cite{kibble,damour,olmez1,olmez2} might produce a detectable SGWB at this level as well.
Similar sensitivity advances will also be made with the directional searches as Advanced LIGO and Advanced Virgo reach their target sensitivities. In fact, the addition of Advanced Virgo to the network, with its long distance displacement from the LIGO sites, will make a further important contribution to the directional searches and their ability to map the sky~\cite{PhysRevLett.118.121102}. This will also be true for the addition of KAGRA~\cite{PhysRevD.88.043007} and LIGO-India~\cite{doi:10.1142/S0218271813410101} to the global network. One can expect to see many important results pertaining to the search for a SGWB from LIGO, Virgo and KAGRA in the coming years.

\section*{Acknowledgments}
Thanks to Tania Regimbau for comments on the manuscript.
NLC is funded by NSF grants PHY-1505373 and PHY-1806990.
The authors gratefully acknowledge the support of the United States
National Science Foundation (NSF) for the construction and operation of the
LIGO Laboratory and Advanced LIGO as well as the Science and Technology Facilities Council (STFC) of the
United Kingdom, the Max-Planck-Society (MPS), and the State of
Niedersachsen/Germany for support of the construction of Advanced LIGO 
and construction and operation of the GEO600 detector. 
Additional support for Advanced LIGO was provided by the Australian Research Council.
The authors gratefully acknowledge the Italian Istituto Nazionale di Fisica Nucleare (INFN),  
the French Centre National de la Recherche Scientifique (CNRS) and
the Foundation for Fundamental Research on Matter supported by the Netherlands Organisation for Scientific Research, 
for the construction and operation of the Virgo detector
and the creation and support  of the EGO consortium. 
The authors also gratefully acknowledge research support from these agencies as well as by 
the Council of Scientific and Industrial Research of India, 
the Department of Science and Technology, India,
the Science \& Engineering Research Board (SERB), India,
the Ministry of Human Resource Development, India,
the Spanish  Agencia Estatal de Investigaci\'on,
the Vicepresid\`encia i Conselleria d'Innovaci\'o, Recerca i Turisme and the Conselleria d'Educaci\'o i Universitat del Govern de les Illes Balears,
the Conselleria d'Educaci\'o, Investigaci\'o, Cultura i Esport de la Generalitat Valenciana,
the National Science Centre of Poland,
the Swiss National Science Foundation (SNSF),
the Russian Foundation for Basic Research, 
the Russian Science Foundation,
the European Commission,
the European Regional Development Funds (ERDF),
the Royal Society, 
the Scottish Funding Council, 
the Scottish Universities Physics Alliance, 
the Hungarian Scientific Research Fund (OTKA),
the Lyon Institute of Origins (LIO),
the Paris \^{I}le-de-France Region, 
the National Research, Development and Innovation Office Hungary (NKFIH), 
the National Research Foundation of Korea,
Industry Canada and the Province of Ontario through the Ministry of Economic Development and Innovation, 
the Natural Science and Engineering Research Council Canada,
the Canadian Institute for Advanced Research,
the Brazilian Ministry of Science, Technology, Innovations, and Communications,
the International Center for Theoretical Physics South American Institute for Fundamental Research (ICTP-SAIFR), 
the Research Grants Council of Hong Kong,
the National Natural Science Foundation of China (NSFC),
the Leverhulme Trust, 
the Research Corporation, 
the Ministry of Science and Technology (MOST), Taiwan
and
the Kavli Foundation.
The authors gratefully acknowledge the support of the NSF, STFC, MPS, INFN, CNRS and the
State of Niedersachsen/Germany for provision of computational resources.%
This article has been assigned the document number LIGO-P1900145.

\section*{References}
\bibliography{moriond}

\end{document}